\documentclass[preprint,12pt]{elsarticle}
\usepackage{varioref}
\usepackage{hyperref}
\usepackage{amsmath}					
\usepackage{amssymb} 									
\usepackage{gensymb}
\usepackage{upgreek}
\usepackage{xcolor}
\usepackage{multirow}
\usepackage{xfrac}

\usepackage{amssymb}
\begin{document}

\begin{frontmatter}

%% Title, authors and addresses

%% use the tnoteref command within \title for footnotes;
%% use the tnotetext command for theassociated footnote;
%% use the fnref command within \author or \address for footnotes;
%% use the fntext command for theassociated footnote;
%% use the corref command within \author for corresponding author footnotes;
%% use the cortext command for theassociated footnote;
%% use the ead command for the email address,
%% and the form \ead[url] for the home page:
%% \title{Title\tnoteref{label1}}
%% \tnotetext[label1]{}
%% \author{Name\corref{cor1}\fnref{label2}}
%% \ead{email address}
%% \ead[url]{home page}
%% \fntext[label2]{}
%% \cortext[cor1]{}
%% \address{Address\fnref{label3}}
%% \fntext[label3]{}

\title{Damage dose dependence of deuterium retention in high-temperature self-ion irradiated tungsten}

%% use optional labels to link authors explicitly to addresses:
%% \author[label1,label2]{}
%% \address[label1]{}
%% \address[label2]{}

\author[ipp]{M.~Zibrov\corref{cor1}}\ead{Mikhail.Zibrov@ipp.mpg.de}
\author[ipp]{T.~Schwarz-Selinger}
\author[kit]{M.~Klimenkov}
\author[kit]{U.~Jäntsch}
\cortext[cor1]{Corresponding author}
\address[ipp]{Max Planck Institute for Plasma Physics, 85748 Garching, Germany}
\address[kit]{Karlsruhe Institute of Technology (KIT), Institute for Applied Materials - Applied Materials Physics, 76021 Karlsruhe, Germany}

\begin{abstract}
Recrystallized tungsten (W) samples were irradiated by 20~MeV self-ions at 1350~K to peak damage doses in the range of 0.001--2.3~dpa. The irradiation-induced defects were then decorated with deuterium (D) by a gentle D plasma exposure ($<5$~eV/D, $5.6 \times 10^{19}$~$\text{D} / (\text{m}^2 \text{s})$) at 370~K. The D depth profiles in the samples were measured using $\rm D(^{3}He,p)\alpha$ nuclear reaction analysis. The maximum trapped D concentration evolves differently with the damage dose compared with the previously studied irradiations at 290~K and 800~K. At the damage doses below 0.1~dpa, the D concentrations are lower than those after the irradiation at 800~K. At higher damage doses, the D concentrations exceed the 800~K values and reach 1.7~at.\% at 2.3~dpa, showing no clear tendency towards saturation. Transmission electron microscopy revealed the presence of nm-sized voids in the samples irradiated at 1350 K, in contrast to the ones irradiated at 290~K and 800~K. Thermal desorption spectroscopy (TDS) indicates that the dominant D trapping sites are different compared to the irradiations at 290~K and 800~K.  Reaction-diffusion simulations show that the TDS spectra can be described by assuming that D is trapped as $\rm D_2$ gas in the void volume and as D atoms at the void surface.
\end{abstract}


\end{frontmatter}

%% \linenumbers

%% main text
\section{Introduction}
Tungsten (W) is the baseline plasma-facing material for ITER and DEMO for several reasons, one being the small solubility of tritium in the pristine lattice. During deuterium-tritium plasma operation, W components will be irradiated by 14.1~MeV neutrons. Neutron bombardment creates displacement damage in the lattice, causes material transmutation and production of hydrogen and helium \cite{Federici_2017,Ueda_2017}. The W components in DEMO will operate at elevated temperatures (673--1300~K) \cite{Zinkle2022}, which will affect their microstructural evolution under irradiation \cite{ZINKLE202091,BHATTACHARYA2020406,HU2022153856}. The neutron-induced defects will deteriorate the thermo-mechanical properties of W. In addition, they will act as trapping sites for hydrogen isotopes, which can significantly increase the tritium inventory in the W components.

Due to the lack of a high-flux 14.1~MeV neutron source, various aspects of neutron irradiation are often simulated using fission neutron or ion beam irradiation. MeV self-ion irradiation is frequently used for simulating the neutron-induced displacement damage since it produces a similar recoil spectrum and does not alter the material composition \cite{WAS2020468}. It also allows achieving high damage doses in relatively short times, though the extrapolation to fusion neutron irradiation requires consideration of the dose rate effects \cite{ZINKLE202091,BHATTACHARYA2020406,WAS2020468,D5MA00677E,ZHAO2025101956}. 

MeV heavy ion irradiation has been widely used for studying the influence of the displacement damage on deuterium (D) trapping in W. Most of the existing studies are limited to irradiation near room temperature (see the review \cite{Schwarz-Selinger_2023}), when the irradiation-induced vacancies in W are immobile. In this case the trapped D concentration reaches saturation at a damage dose near 0.1 displacement per atom (dpa). The corresponding saturation of the density of vacancy-type defects (single vacancies and small vacancy clusters) is confirmed by molecular dynamics simulations \cite{PhysRevMaterials.5.095403,Boleininger2023}, positron annihilation spectroscopy measurements \cite{Hollingsworth2022,WIELUNSKAKUS2024101610}, and thermal diffusivity measurements \cite{Reza2020}. 

Above 600~K the irradiation-induced vacancies in W are mobile and can agglomerate into clusters, eventually forming nm-sized voids \cite{HU2022153856,KLIMENKOV2022154018,KLIMENKOV2024154950,KLIMENKOV2025155673,WANG2024119942,WANG2026122148}. Such voids are expected to have a different D trapping mechanism than single vacancies and small vacancy clusters \cite{Hou2019,Ren2019}. Namely, a considerable part of the trapped D can be in the form of $\rm D_2$ gas in the void volume \cite{VANVEEN19881113,bubble1,bubble2}. In addition, the damage dose dependence of the defect density at high temperatures can differ significantly from that at low temperatures, particularly in the void swelling regime \cite{ZINKLE202091,BHATTACHARYA2020406,ZHAO2025101956,KLIMENKOV2025155673}. 

There is a limited number of studies on D retention in high-temperature ion-irradiated (up to 1273~K) \cite{OGORODNIKOVA2014379,SAKURADA2016141,Simmonds2017,Markelj_2019,KOBAYASHI20191624,WANG2021152749,Markelj_2022,ZHU2024101620} and fission-neutron-irradiated W (up to 1379~K), see \cite{OYA2021100980} and references therein. Except for a few studies where the ion irradiation and the D exposure occurred simultaneously \cite{Markelj_2019,Markelj_2022}, the D exposure is often performed at a fixed relatively low temperature to decouple the effects of the defect annealing and the D thermal detrapping. It was observed that (at a fixed damage dose) the trapped D concentration monotonically decreased with increasing irradiation temperature due to defect annealing. Our recent study \cite{tms-800K-data} showed that the damage dose dependence of D retention in W after self-ion irradiation at 800~K is similar to that at 290~K. Also the shapes of the thermal desorption spectra from the samples irradiated at 290~K and 800~K were very similar. Only the absolute concentrations of trapped D ware a factor of four lower, suggesting similar defect structure with reduced density. No voids were detected in these samples \cite{CHROMINSKI20191}. In another recent study we found unexpectedly high trapped D concentration (0.9~at.\%) in W irradiated at 1350~K to 0.3~dpa \cite{W-Re-paper-Zibrov}. The TDS spectrum from this sample looked significantly different compared with that after irradiation at 290~K, suggesting a significant change of the defect structure.

In this contribution we present the measurements of the damage dose dependence (0.001--2.3~dpa) of the D retention in W after self-ion irradiation at 1350~K. These are accompanied by transmission electron microscopy investigations, confirming the presence of nm-sized voids in the samples. First reaction-diffusion simulations show that the TDS spectra can be described by assuming that D is trapped as $\rm D_2$ gas in the void volume and as D atoms at the void surface.

\section{Experimental details}

\subsection{Sample preparation}
$12 \times 15 \times 0.8$~$\text{mm}^3$ samples made of hot-rolled polycrystalline W with a purity of 99.97~wt.\,\% (Plansee SE) were used. The samples were consecutively ground with SiC sandpapers with decreasing grit sizes (up to P4000). Then they were electrochemically polished to a mirror finish in a 1.5\,\% $\rm NaOH$ solution \cite{Manhard2013}. Afterwards they were ultrasonically cleaned with acetone and high purity ethanol. Later they were recrystallized at 2000~K for 2~min at a pressure below $10^{-6}$~Pa. As a result, the material exhibits 10--50~$\mu$m large grains and a low dislocation density of $2 \times 10^{12}$~$\text{m} / \text{m}^{3}$ \cite{manhard-metall-2015}.

\subsection{Ion irradiation}
The samples were irradiated by 20~MeV W$^{6+}$ ions in the implantation chamber described in \cite{SCHWARZSELINGER2017683}. A sample was pressed against a boron nitrite coated resistive heater (Boralectric{\textregistered}) by a molybdenum mask. A thin tantalum sheet was placed in between the heater and the sample to avoid contamination. Irradiations were performed at a sample temperature of 1350~K, as measured by two type K thermocouples. One was clamped to the sample surface, the other one to the resistive heater. The ion flux was continuously measured during the course of the irradiation using Faraday cups located at the four corners of a copper diaphragm placed in front of the sample holder. A focussed ion beam (2~mm FWHM) was raster scanned (1~kHz) across a square area including the Faraday cups to assure a laterally homogeneous irradiation. From the size of the Faraday cups and the measured current over time, the acquired ion fluence was determined. The diaphragm was cooled with liquid nitrogen to reduce carbon contamination of the sample surface during the irradiation \cite{WAS201758}. The samples were placed off-center to the hole in the diaphragm, such that approximately \sfrac{2}{3} of the sample surface was irradiated and \sfrac{1}{3} remained unaffected. Nuclear reaction analysis (NRA) using the $\rm ^{12}C(^{3}He,p_{0,1,2})^{14}N$ \cite{HESS2024165141} and $\rm ^{16}O(^{3}He,p_0)^{18}F$ \cite{GUITARTCOROMINAS201913} reactions showed that the irradiated and unirradiated samples have similar small amounts of carbon and oxygen on the surface below $10^{20}$~$\text{at.} / \text{m}^{2}$.

The depth distribution of the primary radiation damage was calculated using the ``quick calculation of damage'' mode in SRIM-2008 \cite{srim} with the displacement threshold energy of 90~eV \cite{astm} and the lattice binding energy of 0~eV \cite{STOLLER201375}. The peak damage doses corresponding to the used fluences ($3.15\times 10^{15}-7.87\times 10^{18}$~$\text{W-ions} / \text{m}^2$) are in the range of 0.001--2.3~dpa. The average damage dose rates for the different samples were in the range of $(2.0-3.5) \times 10^{-5}$~$\text{dpa} / \text{s}$. The peak damage dose rates were two orders of magnitude larger.

\subsection{Deuterium plasma exposure}\label{sec:plasma-exp-descr}
To decorate the irradiation-induced defects with D, the samples were exposed to a D plasma in the PlaQ device \cite{0963-0252-20-1-015010}. Six samples were screwed to a W-coated copper holder at the same radial distance from its center. The holder was kept at 370~K during the exposure. The total D ion flux was $5.6 \times 10^{19}$~$\text{D} / (\text{m}^2 \text{s})$ and was dominated by $\rm D_3^+$ ions. The holder was electrically floating, which leads to an ion energy of $<15$~eV for the majority ion species $\rm D_3^+$ and, hence, $<5$~eV/D. The samples were first exposed to a fluence of $2\times 10^{25}$~$\text{D} / \text{m}^2$. After performing the \textit{ex situ} ion beam analysis, the samples were exposed again to the same D fluence, corresponding to a total accumulated fluence of $4\times 10^{25}$~$\text{D} / \text{m}^2$. Scanning electron microscopy investigations showed no signs of blister formation on the plasma-exposed surfaces.

\subsection{Deuterium retention measurements}
NRA measurements utilizing the $\rm D(^{3}He,p)\alpha$ reaction were carried out using eight different $\rm ^{3}He$ energies ranging from 0.5~MeV to 4.5~MeV \cite{Mayer2009506}. Protons were detected under $135\degree$ and $175\degree$, and $\alpha$-particles under $102\degree$. Measurements were conducted with a fixed nominal collected ion dose of 10 to 20~$\mu C$ on a beam spot of $1 \times 1$~$\text{mm}^2$ for each energy. The most probable D concentration profile corresponding to the measured $\rm p$ and $\alpha$ energy spectra was determined using NRADC \cite{Schmid201264} coupled with SIMNRA 7.04 \cite{mayer-simnra}. The collected ion dose was cross-checked with backscattered $\rm ^{3}He$ detected under $165\degree$. The reaction cross-section data was taken from Wielunska et al.~\cite{WIELUNSKA201641} and Möller and Besenbacher \cite{MOLLER1980111}. $^{3}\rm He-NRA$ depth profiling after the second D loading was performed on a different location on the sample to exclude interference with the $^{3}\rm He$ implanted during the first depth profiling.

Thermal desorption spectroscopy (TDS) measurements were performed at the TESS setup \cite{SCHWARZSELINGER2025102029}. A sample was placed inside a quartz tube attached to a vacuum vessel and was heated by a tubular furnace. The furnace was linearly heated with a rate of 0.05~K/s up to 1000~K. The corresponding sample temperature response was measured in independent experiments using a thermocouple spot-welded to one of the used samples. 16 mass channels between 1 and 44 amu/q were followed with a high-end quadrupole mass-spectrometer (QMS, Balzers DMM 422) as a function of time during the temperature ramp. The QMS signal of $\rm D_2$ molecules was quantified after each measurement using a calibrated leak bottle. The calibration factors for $\rm HD$, $\rm HDO$, and $\rm D_2O$ molecules were previously determined to be 0.7 of that for $\rm D_2$ molecules \cite{SCHWARZSELINGER2025102029}. The TDS spectra shown in the paper are the sum of $\rm D_2$, $\rm HD$, $\rm HDO$, and $\rm D_2O$ desorption fluxes. The contribution of $\rm D_2$ molecules to the total amount of released D was ranging from 65\% for the sample irradiated to 0.005~dpa to 88\% for the sample irradiated to 2.3~dpa.

\subsection{Transmission electron microscopy}
Following the TDS measurements, the samples irradiated to 0.1, 0.5, and 2.3~dpa were investigated using transmission electron microscopy (TEM). Since the irradiations were carried out at 1350~K, it is expected that the sample heating up to 1000~K during TDS did not considerably alter the material microstructure \cite{Wang_2026}. A TEM lamella was cut perpendicularly to the irradiated surface using a $\rm Ga^+$ focused ion beam (FIB). It was taken from a surface region not subjected to the ion beam analysis. The lamellae were flash electropolished in a 1\,\% NaOH solution to remove the FIB artifacts \cite{HORVATH201929}. Investigations were carried out using a Talos F200X TEM. 
The voids were imaged in a bright field mode using underfocus conditions, which makes the voids appear bright in comparison to the matrix. The ImageJ software was used for the statistical analysis of the TEM images. The lamella thickness was determined using electron energy loss spectroscopy. Due to the local thickness variations caused by the flash polishing, the average thickness of the examined area was used in the defect density calculations. The void swelling was calculated as
\begin{equation}
\text{Void swelling (\%)} = 100 \frac{\frac{\pi}{6}  \sum_{i=1}^{N} d_i^3}{lwt-\frac{\pi}{6}  \sum_{i=1}^{N} d_i^3},
\end{equation}
where $d_i$ is the individual void diameter, $N$ is the number of counted voids, $l$ and $w$ are the length and width of the investigated region, and $t$ is the average lamella thickness.

\section{Results and discussion}

\subsection{Microstructure of the irradiated samples}\label{sec:TEM-results}
\begin{figure}[!htb]
\center{\includegraphics[width=1.0\textwidth]{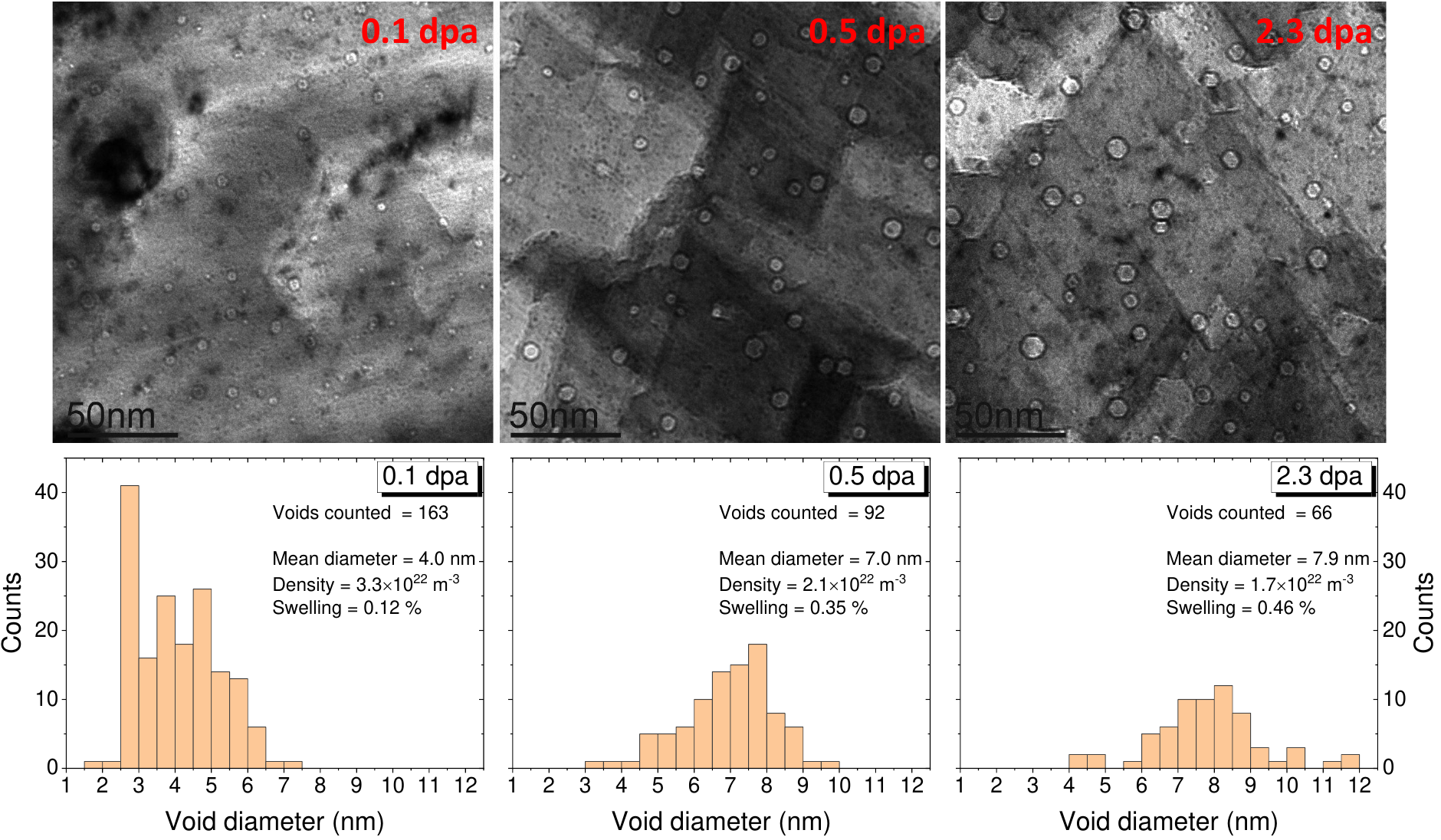}}
\caption{TEM images and the corresponding void size distribution histograms for W irradiated with 20~MeV W ions at 1350~K to the peak damage doses of 0.1~dpa (left column), 0.5~dpa (middle column), and 2.3~dpa (right column). The background structures in the TEM images are artefacts of the flash polishing.}
\label{fig:TEM-dpa-series}
\end{figure}

\begin{figure}[!htb]
\center{\includegraphics[width=1.0\textwidth]{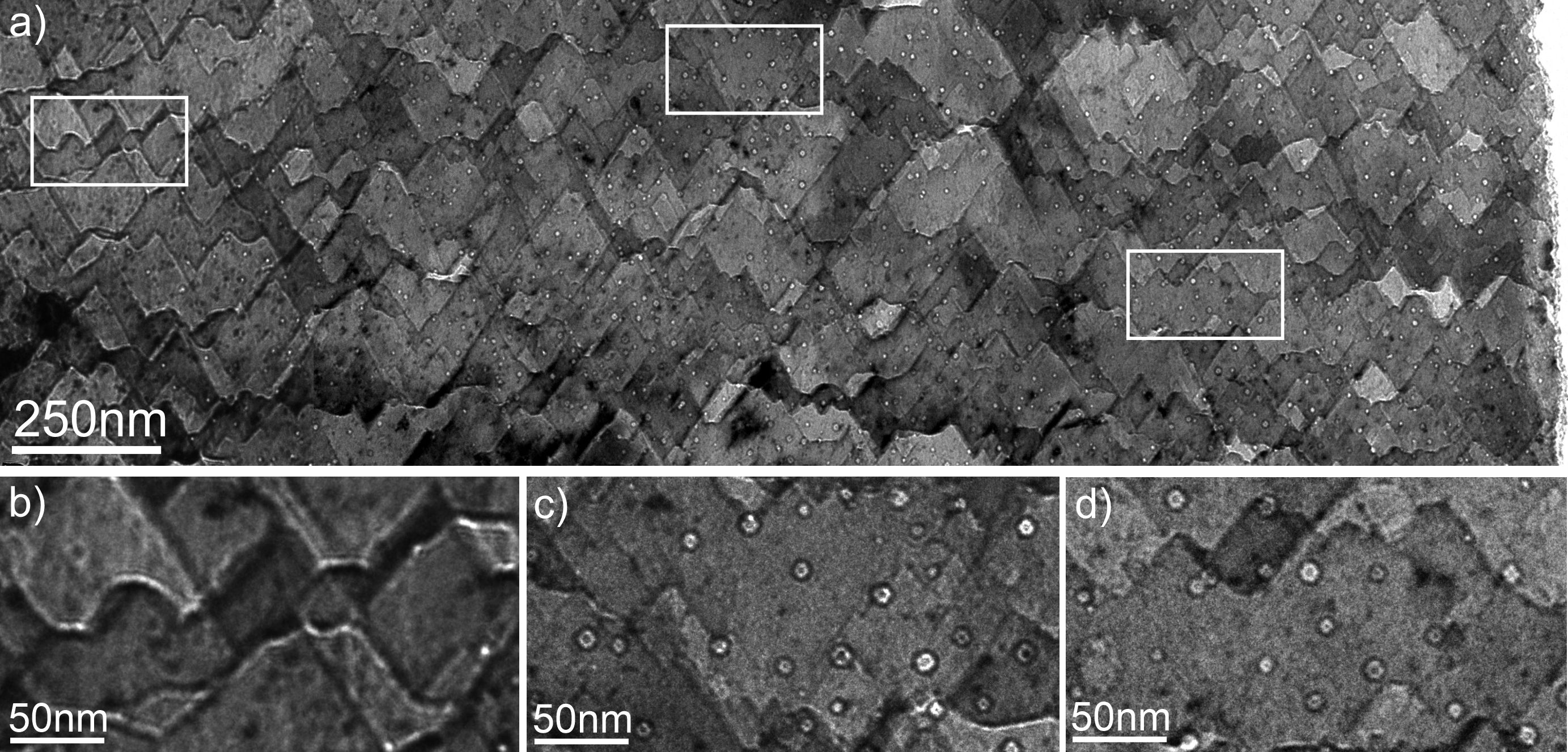}}
\caption{(a) TEM image showing the void distribution over the sample thickness in W irradiated with 20~MeV W ions at 1350~K to a peak damage dose of 2.3~dpa. The sample surface is on the right. The parts (b-d) show the images from the areas marked by rectangles above in (a). The background structures in the TEM images are artefacts of the flash polishing.}
\label{fig:TEM-depth-res}
\end{figure}

TEM investigations detected the presence of nm-sized voids in the studied samples (0.1, 0.5, and 2.3~dpa), as illustrated in Fig.~\ref{fig:TEM-dpa-series}. The visible background structures in the TEM images are due to the fluctuations of the lamellae thickness, which are caused by the flash polishing. Fig.~\ref{fig:TEM-depth-res} shows the depth distribution of voids in the sample irradiated to 2.3~dpa. The voids shown in the parts (c) and (d), which are located at depths of 1.4~$\mu$m and 0.6~$\mu$m, respectively, exhibit similar sizes and number density. Beyond a depth of about 2.6~$\mu$m, the voids are no longer present (b). This agrees with the thickness of the primary damage zone calculated in SRIM (see Fig.~\ref{ris:NRA-D-flc-var}). There is no apparent void-denuded zone next to the surface, which is frequently observed in other metals irradiated at elevated temperatures \cite{SHAO2025113553}.  

The histograms of the void size distribution are shown in Fig.~\ref{fig:TEM-dpa-series}. The statistical analysis of the TEM data is listed in Table~\ref{tab:TEM-stats}. The average void size increases with increasing damage dose, while the void number density decreases. A similar tendency has been observed in fission neutron-irradiated W \cite{KLIMENKOV2024154950,KLIMENKOV2025155673}. The void swelling (reflecting the total void volume fraction in the material) increases with increasing damage dose. Overall, the void sizes, number densities, and swelling found in our samples are of the same magnitude as those found in neutron-irradiated W \cite{HU2022153856,KLIMENKOV2022154018,KLIMENKOV2024154950,KLIMENKOV2025155673}.

\begin{table}
\caption{Average void diameter, number density, and void swelling in W irradiated with 20~MeV W ions at 1350~K to various peak damage doses.}
\label{tab:TEM-stats}
\begin{center}
\begin{tabular}{|p{0.23\textwidth}|p{0.23\textwidth}|p{0.23\textwidth}|p{0.23\textwidth}|}
\hline
Peak damage dose (dpa) & Average void diameter (nm) & Void number density ($1/ \text{m}^{3}$) & Void swelling (\%)  \\
\hline
0.1 & 4.0 & $3.3 \times 10^{22}$ & $0.12\pm 0.02$ \\
0.5 & 7.0 & $2.1 \times 10^{22}$ & $0.35\pm 0.04$ \\
2.3 & 7.9 & $1.7 \times 10^{22}$ & $0.46\pm 0.04$ \\
\hline
\end{tabular}
\end{center}
\end{table}

\subsection{Deuterium retention}
\begin{figure}[!htb]
\center{\includegraphics[width=0.5\textwidth]{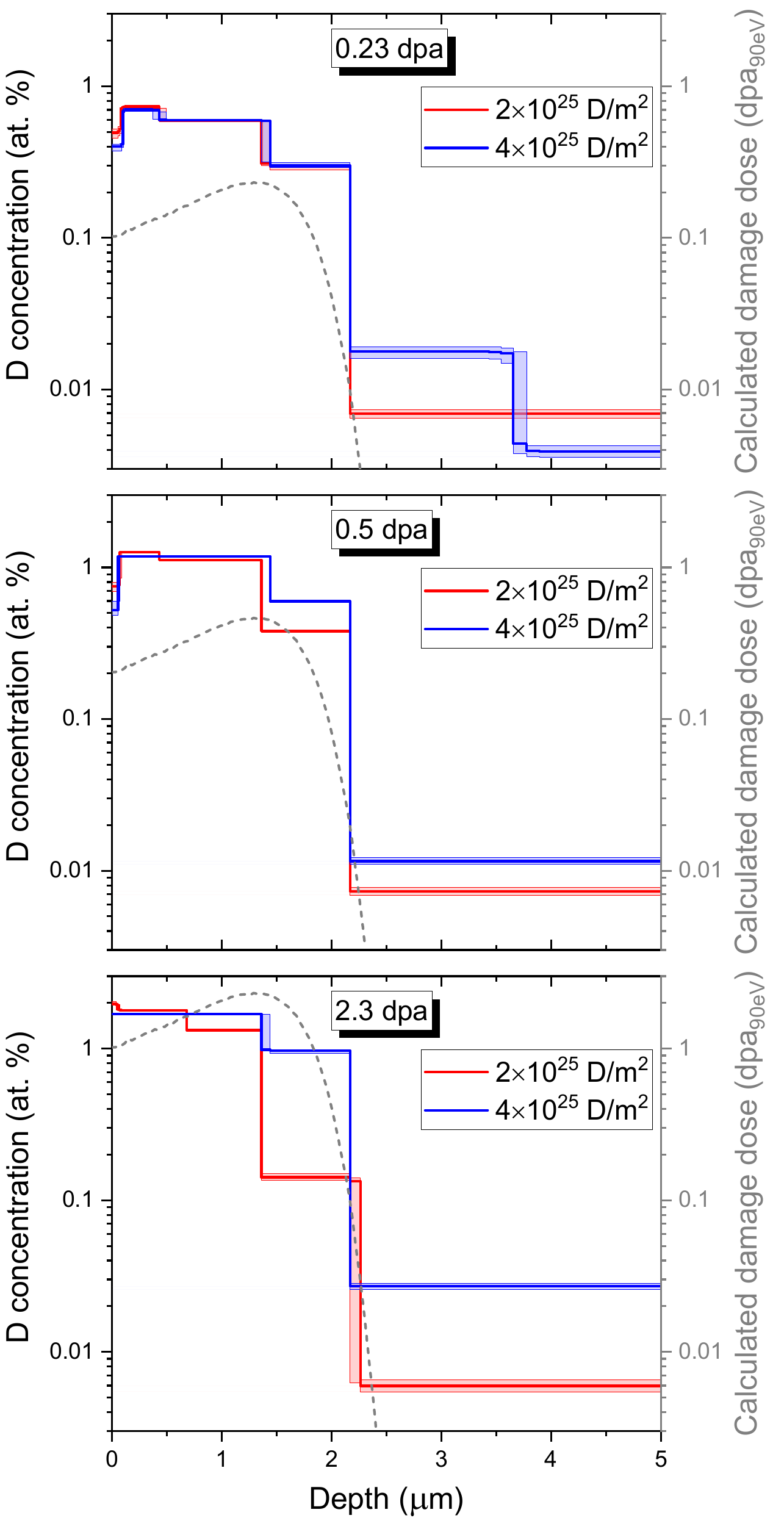}}
\caption{D concentration profiles in W irradiated with 20~MeV W ions at 1350~K to the peak damage doses of 0.23~dpa (upper panel) 0.5~dpa (middle panel) and 2.3~dpa (lower panel). The samples were exposed to a D plasma with a mean ion energy $<5$~eV/D to the fluences of $2\times 10^{25}$~$\text{D} / \text{m}^2$ and $4\times 10^{25}$~$\text{D} / \text{m}^2$ at a sample temperature of 370~K. The grey dashed lines show the corresponding primary damage profiles calculated using SRIM.}
\label{ris:NRA-D-flc-var}
\end{figure}

Fig.~\ref{ris:NRA-D-flc-var} shows the D depth profiles for two D fluences and three different damage doses. In the case of the samples irradiated to the peak damage doses up to $0.23$~dpa, there is a marginal difference in the D depth profiles (and D retention) for the fluences of $2\times 10^{25}$~$\text{D} / \text{m}^2$ and  $4\times 10^{25}$~$\text{D} / \text{m}^2$, as shown for the $0.23$~dpa case in Fig.~\ref{ris:NRA-D-flc-var}. This confirms the complete decoration of the irradiation-induced defects with D already after the first D decoration fluence. Concurrently, in the case of the samples irradiated to 0.5~dpa and 2.3~dpa (Fig.~\ref{ris:NRA-D-flc-var}), D penetrates deeper with increasing fluence, while the maximum D concentration almost does not change. Later study has confirmed that the fluence of $4\times 10^{25}$~$\text{D} / \text{m}^2$ was sufficient to decorate all irradiation-induced defects with D in the sample irradiated to 2.3~dpa \cite{Agnoli-thesis}. The width of the zone with high D concentration agrees with the width of the void-containing zone observed by TEM and with the SRIM calculation.

\begin{figure}[!htb]
\center{\includegraphics[width=0.6\textwidth]{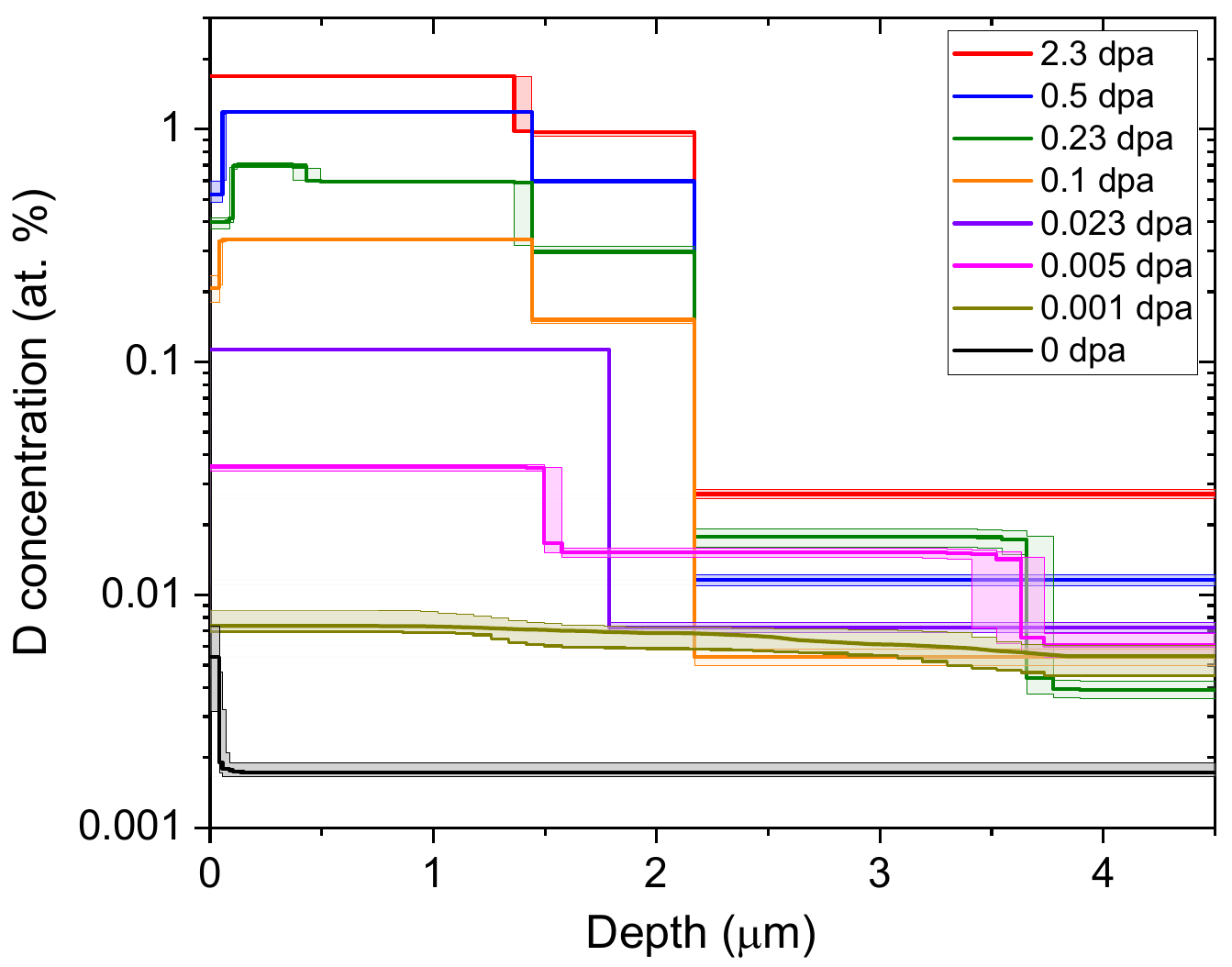}}
\caption{D concentration profiles in W irradiated with 20~MeV W ions at 1350~K to various peak damage doses. The samples were exposed to a D plasma with a mean ion energy $<5$~eV/D to the fluence of $4\times 10^{25}$~$\text{D} / \text{m}^2$ at a sample temperature of 370~K. The 0~dpa and 0.001~dpa samples were exposed to the fluence of $2\times 10^{25}$~$\text{D} / \text{m}^2$.} 
\label{ris:NRA-all-doses}
\end{figure}

\begin{figure}[!htb]
\center{\includegraphics[width=0.6\textwidth]{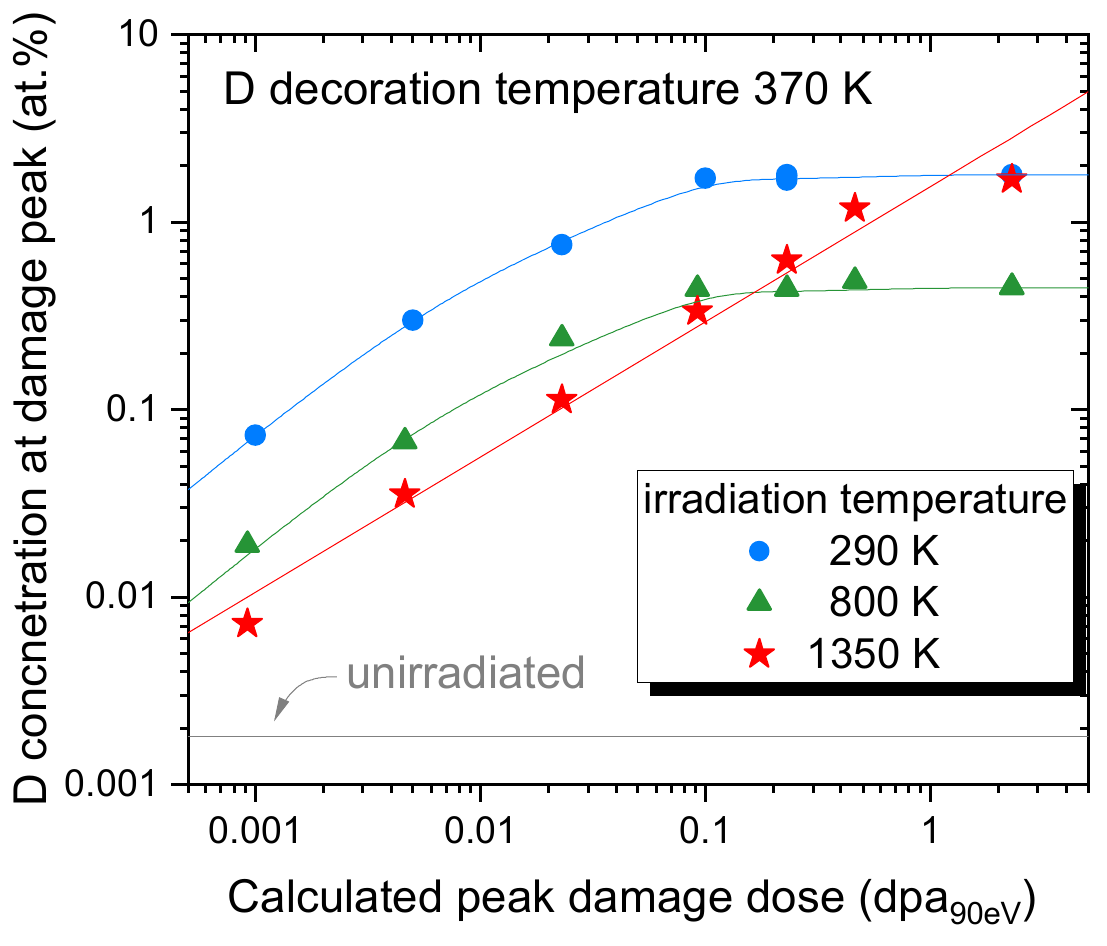}}
\caption{Maximum trapped D concentration as a function of the peak damage dose in W irradiated with 20~MeV W ions at 1350~K (stars). The previous data for the irradiations at 290~K (circles) and 800~K (triangles) is shown for the comparison \cite{tms-800K-data}. The samples were exposed to a D plasma with a mean ion energy $<5$~eV/D at a sample temperature of 370~K. The lines are only a guide for the eye.}
\label{ris:D-max-vs-dose}
\end{figure}

Fig.~\ref{ris:NRA-all-doses} shows the D concentration profiles in the samples irradiated to various peak damage doses after the D plasma exposure to the fluence of $4\times 10^{25}$~$\text{D} / \text{m}^2$. The maximum trapped D concentration increases with increasing peak damage dose, as summarized in Fig.~\ref{ris:D-max-vs-dose}. Because of the same D loading temperature, a quantitative comparison with the previous data where irradiations were conducted at 290~K and at 800~K can be made and is also shown in Fig.~\ref{ris:D-max-vs-dose}. The damage dose dependence for the 1350~K irradiation is clearly different compared with the cases of irradiation at 290~K and 800~K \cite{tms-800K-data}. At damage doses up to 0.1~dpa, the trapped D concentration in the samples irradiated at 1350~K is smaller compared with that in the samples irradiated at 290~K and 800~K. This can be attributed to the increased defect annealing rate\footnote{Note that irradiation to the lowest dose of 0.001~dpa takes only a few minutes, while cooling down from 1350~K to 600~K takes half an hour. Therefore, defect annealing during the cool-down phase may be non-negligible for the low-dose samples.}. While in the case of the samples irradiated at 290~K and 800~K the trapped D concentration reaches saturation at damage doses above 0.1~dpa, there is no clear trend  in the case of irradiation at 1350~K. At the highest damage dose of 2.3~dpa the trapped D concentration almost coincides with the saturation D concentration in the samples irradiated at 290~K. The D concentration in the 2.3~dpa sample is about five times greater than that in the 0.1~dpa sample. This correlates with the fourfold increase of the void swelling (Section~\ref{sec:TEM-results}). Hence, the voids likely play a significant role in D retention in these samples.

\begin{figure}[!htb]
\center{\includegraphics[width=0.6\textwidth]{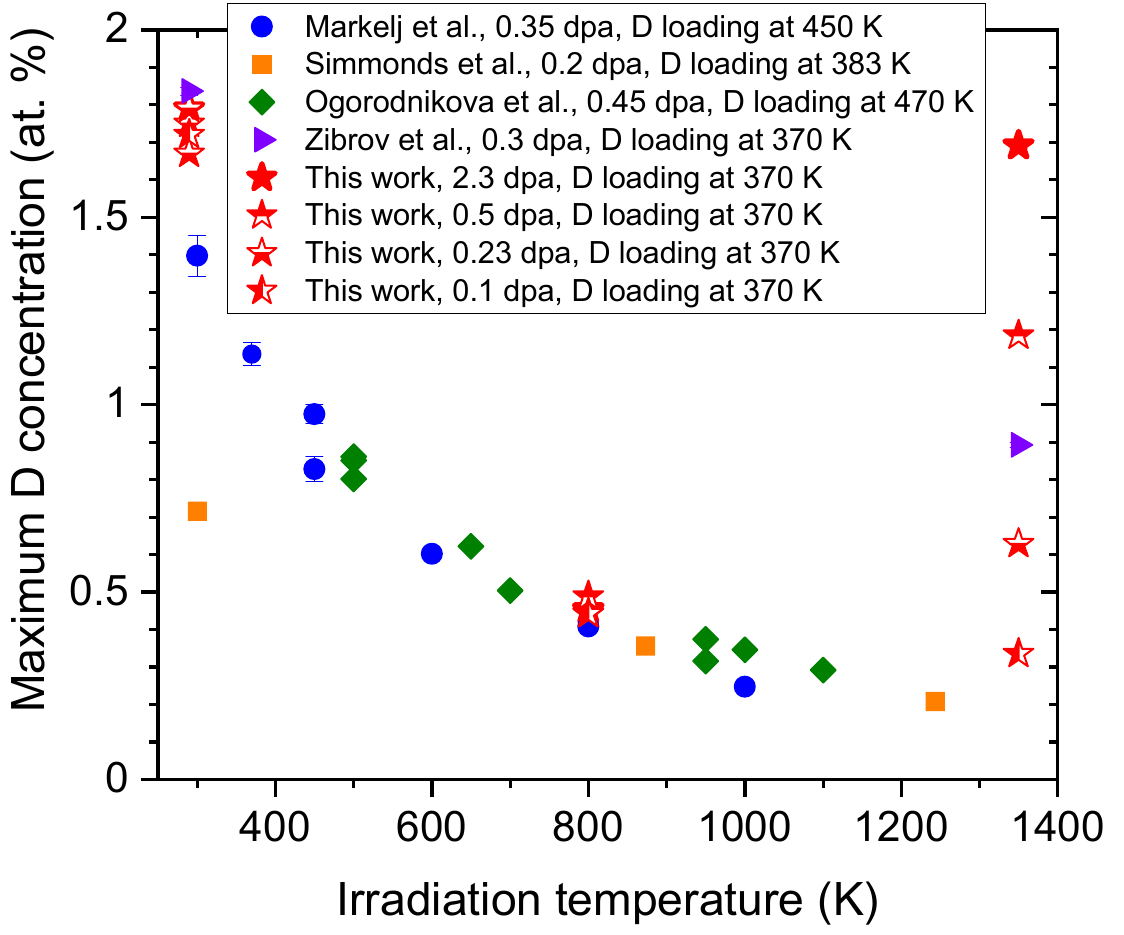}}
\caption{Maximum trapped D concentration as a function of the ion irradiation temperature reported by Ogorodnikova et al.~\cite{OGORODNIKOVA2014379}, Simmonds et al.~\cite{Simmonds2017}, Markelj et al.~\cite{Markelj_2022} and Zibrov et al.~\cite{W-Re-paper-Zibrov}. The present data for different peak damage doses is also shown.}
\label{ris:Lit-data-T-dep}
\end{figure}

The maximum D concentrations in the samples irradiated at 1350~K can also be compared with the previously reported D concentrations in the samples irradiated to 0.2--0.5~dpa at various temperatures and then loaded with D at 370--470~K (Fig.~\ref{ris:Lit-data-T-dep}). In the case of irradiation at temperatures up to 1243~K, there is a monotonic decrease of the D concentration with increasing irradiation temperature. However, the D concentrations in the samples irradiated at 1350~K to 0.1--2.3~dpa are larger than expected from a simple extrapolation of the literature data.  
The void swelling in metals usually has a bell-shaped temperature dependence \cite{BHATTACHARYA2020406}. The maximum void swelling of 0.9~\% in W neutron-irradiated to 1~dpa was observed at 1373~K \cite{KLIMENKOV2022154018}. The maximum void swelling of 0.21~\% in W irradiated by 6~MeV copper ions at room temperature to 0.6~dpa was observed after post-irradiation annealing at 1273~K \cite{WANG2024119942}. This may explain the unexpectedly high D concentrations and different damage dose dependence observed in the present study. 

\begin{figure}[!htb]
\center{\includegraphics[width=0.6\textwidth]{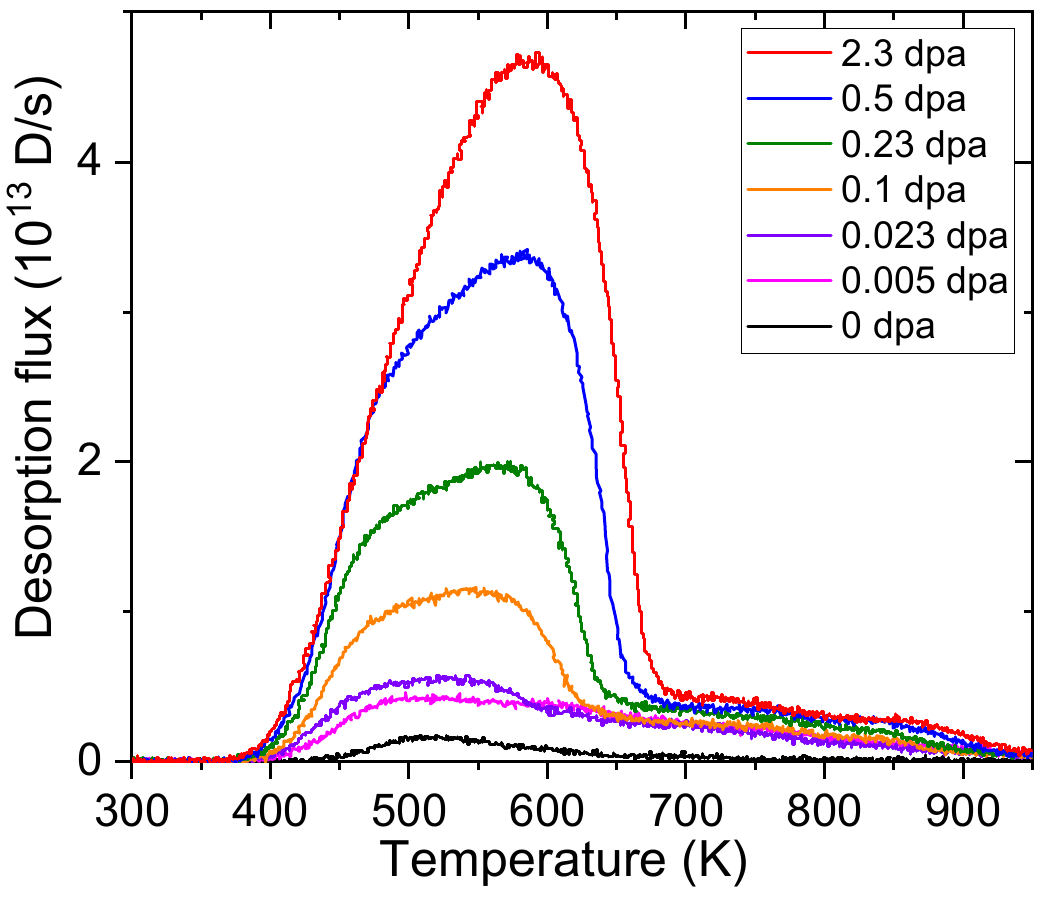}}
\caption{TDS spectra from W irradiated with 20~MeV W ions at 1350~K to various peak damage doses. The samples were exposed to a D plasma with a mean ion energy $<5$~eV/D to the fluence of $4\times 10^{25}$~$\text{D} / \text{m}^2$ at a sample temperature of 370~K. The 0~dpa sample was exposed to the fluence of $2\times 10^{25}$~$\text{D} / \text{m}^2$.}
\label{ris:TDS-vs-dose}
\end{figure}

The TDS spectra from the samples irradiated at 1350~K (Fig.~\ref{ris:TDS-vs-dose}) have a different shape compared to those from the samples irradiated at temperatures up to 1000~K \cite{Markelj_2019,tms-800K-data}. The latter show typically a two peak structure. The first broad desorption maximum is reached at around 550~K and the second, slightly smaller structure peaks around 800~K.  The spectra from the present samples irradiated at 1350~K do not have this second prominent peak near 800~K, but rather a long low-intensity tail. This agrees with our previous observation for W single crystals irradiated at 1350~K to 0.3~dpa \cite{W-Re-paper-Zibrov} and points towards a different defect nature. As discussed above, the different damage dose dependence of trapped D concentration (which reflects the defect density) also points to fundamental differences compared to irradiations at lower temperatures. The presence of nm-sized voids in the samples irradiated to 0.1, 0.5, and 2.3~dpa was confirmed by TEM (Section~\ref{sec:TEM-results}). Concurrently, no apparent voids were found in the samples irradiated at 800~K \cite{CHROMINSKI20191}. It is established that irradiation at 290~K results in the creation of mostly single vacancies and small vacancy clusters \cite{PhysRevMaterials.5.095403,Boleininger2023,WIELUNSKAKUS2024101610}. Owing to the similar damage dose dependence and shapes of the TDS spectra \cite{Markelj_2019,tms-800K-data}, it is likely that the defects are similar also in the case of irradiation at 800~K, when vacancies are already mobile. It appears that irradiation at 800~K resulted only in a reduced defect density compared with 290~K irradiation due to higher defect annealing rate. However, formation of voids in W with an average diameter around 1~nm after ion irradiation at 773~K to 0.02--10~dpa has been reported in other studies \cite{WANG2026122148,HU2021153175}. Possible presence of sub-nm sized voids in W irradiated at 800~K to 0.2~dpa was reported in \cite{ZAVASNIK2025115050}. The absence of voids in our samples irradiated at 800~K may be related with the use of ion beam rastering during irradiation, which can significantly affect the void formation \cite{GETTO2015116,GIGAX2015343}. It is also possible that a significant change of the D trapping behavior, i.e., the D retention is dominated by the $\rm D_2$ gas in the void volume, occurs only in  voids above a certain size.

\subsection{Reaction-diffusion simulations}
As follows from Density Functional Theory (DFT) calculations \cite{Hou2019,Ren2019}, in sufficiently large vacancy clusters D can be present not only as D atoms attached to the cluster surface, but also as $\rm D_2$ molecules in the cluster volume. The number of D atoms and $\rm D_2$ molecules in each nm-sized void can be large, allowing to apply a thermodynamic treatment. In the light of these findings, we developed a reaction-diffusion model of D trapping and release from spherical voids in metals \cite{bubble1,bubble2}, which is briefly outlined in the Appendix~\ref{app:model-descr}.

In the simulations we used the void parameters derived from the TEM observations (Table~\ref{tab:void-par-sim}): number density $N_\mathrm{v,TEM}$, the average radius $R$, the average surface area $A$, and the average volume $V$. The depth distribution of the void number density was approximated by a smooth step-like function: $N_\mathrm{v} (x) = N_\mathrm{v,TEM} \left\lbrace 0.5 - 0.5 \mathrm{tanh}\left[\alpha (x - \ell_\mathrm{v}) \right] \right\rbrace$, where  $\ell_\mathrm{v}$ is the width of the void-containing region and $\alpha  = 5$~$\mathrm{\mu m}^{-1}$ represents the steepness of the step. 

We simulated two consecutive D plasma exposures and TDS, reproducing the full temperature history of the samples. Since $\ell_\mathrm{v}$ is much greater than the width of the D implantation zone \cite{Kapser_2018}, the D plasma exposures were simulated using the maximum interstitial D concentration in the implantation zone $C_\mathrm{max}$ as a boundary condition at the plasma-exposed surface $C(x=0,t) = C_\mathrm{max}$. At the end of the plasma exposures, $C_\mathrm{max}$ was quickly reduced to zero. The interstitial D concentration at the back surface of the sample was always zero. The $C_\mathrm{max}$ computed for our D plasma exposure conditions using the D implantation profile in W given by Kapser et al.~\cite{Kapser_2018} is $1.8 \times 10^{-7}$~at.\%. We found that using $C_\mathrm{max} = 1.45 \times 10^{-7}$~at.\% allows to match closely the D retention in the samples measured after the first D plasma exposure (fluence $2\times 10^{25}$~$\text{D} / \text{m}^2$). The $\ell_\mathrm{v}$ was then adjusted in order to match closely the D retention in the samples measured after the second D plasma exposure (cumulative fluence $4\times 10^{25}$~$\text{D} / \text{m}^2$). 

\begin{table}
\caption{Void parameters used in the simulations.}
\label{tab:void-par-sim}
\begin{center}
\begin{tabular}{|p{0.15\textwidth}|p{0.15\textwidth}|p{0.15\textwidth}|p{0.16\textwidth}|p{0.15\textwidth}|p{0.2\textwidth}|}
\hline
Peak damage dose (dpa) & Number density $N_\mathrm{v,TEM}$ ($1/ \text{m}^{3}$) & Diameter $2R$ (nm) & Surface area $A$ ($\text{nm}^2$) & Volume $V$ ($\text{nm}^3$) & Width of void-containing region $\ell_\mathrm{v}$ ($\mu$m) \\
\hline
0.1 & $3.3 \times 10^{22}$ & 4.0 & 201 & 283 & 1.4 \\
0.5 & $2.1 \times 10^{22}$ & 7.0 & 156 & 192 & 1.85 \\
2.3 & $1.7 \times 10^{22}$ & 7.9 & 56 & 43 & 2.5 \\
\hline
\end{tabular}
\end{center}
\end{table}

The D concentration profile and the TDS spectrum from a plasma-exposed unirradiated W could be reproduced using the conventional trapping model (McNabb and Foster) \cite{Takagi} with one trap type homogeneously distributed through the entire sample. Hence, in addition to voids, this trap type was used in the simulations. The corresponding trap parameters are listed in Tab.~\ref{tabl:sim-param} in the Appendix~\ref{app:model-descr}. This trap gave noticeable contribution to TDS only in the case of the sample irradiated to 0.1~dpa.

\begin{figure}[!htb]
\center{\includegraphics[width=0.5\textwidth]{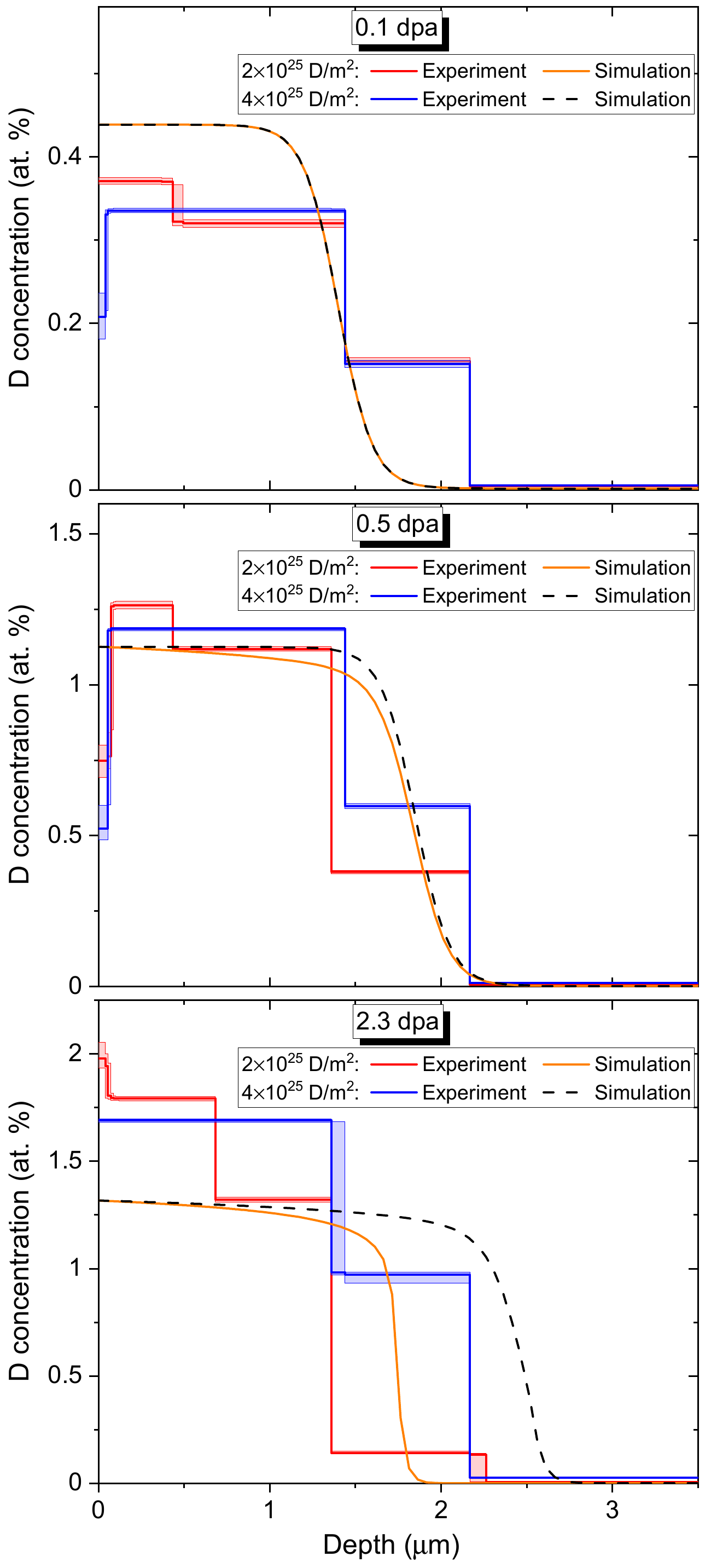}}
\caption{Comparison of experimental and simulated D concentration profiles for the fluences of $2\times 10^{25}$~$\text{D} / \text{m}^2$ and $4\times 10^{25}$~$\text{D} / \text{m}^2$ in the samples irradiated to 0.1~dpa (upper panel), 0.5~dpa (middle panel), and 2.3~dpa (lower panel).}
\label{ris:DP-exp-TESSIM-comp}
\end{figure}

\begin{figure}[!htb]
\center{\includegraphics[width=0.5\textwidth]{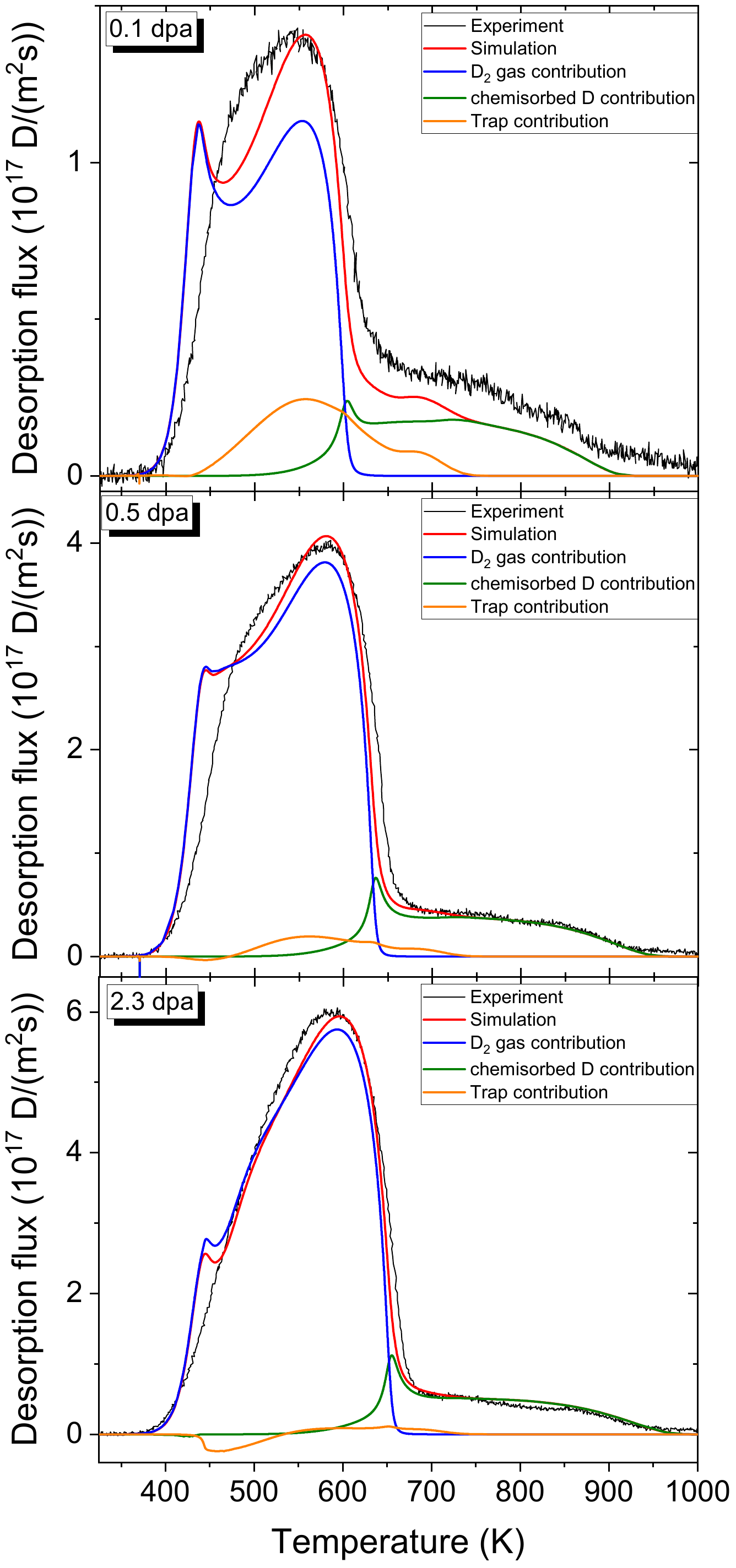}}
\caption{Comparison of experimental and simulated TDS spectra from the samples irradiated to 0.1~dpa (upper panel), 0.5~dpa (middle panel), and 2.3~dpa (lower panel). The contributions of the $\rm D_2$ gas depletion in the void volume, chemisorbed D depletion at the void surface, and the trap in the material bulk to the total TDS spectrum are also shown.}
\label{ris:TDS-exp-sim}
\end{figure}

The simulated TDS peak position depends on the sum $E_\mathrm{S} + E_\mathrm{D}$, while the position of the rising edge of the spectrum depends on the activation energy for the surface-to-bulk transition $E_\mathrm{S} + E_\mathrm{BS} + Q_\mathrm{c}$ \cite{bubble2}. We used the D diffusivity in W reported by Holzner \cite{holzner-phd} with the activation energy $E_\mathrm{D} = 0.265 \pm 0.02$~eV. The TDS peak positions in the simulated spectra would match the experimental ones when using the heat of D solution $E_\mathrm{S}=1.0$~eV, which is lower than $E_\mathrm{S}=1.14 \pm 0.04$~eV reported by Holzner \cite{holzner-phd}. The widely used values reported by Frauenfelder \cite{frauenfelder} are: $E_\mathrm{S} = 1.04 \pm 0.17$~eV and  $E_\mathrm{D} = 0.39 \pm 0.09$~eV. 

We assumed that the long high-temperature tails in the TDS spectra correspond to the release of chemisorbed D at the void surface (after the $\rm D_2$ gas has been depleted from the void volume). It is known from the surface science experiments \cite{ALNOT198929} and DFT calculations \cite{Ajmalghan_2019,Ferro_2023,Hodille_2024} that the heat of chemisorption $Q_\mathrm{c}$ decreases with increasing D surface coverage $\theta = a / a_\mathrm{m}$, where $a$ is the surface concentration of chemisorbed D and $a_\mathrm{m}$ is the corresponding maximum value. We approximated the coverage-dependence by a simple expression: $Q_\mathrm{c} (\theta) = Q_\mathrm{c,0} + \Delta Q_\mathrm{c} \left\lbrace 1 - \exp \left[- \gamma \left(1 - \theta \right) \right] \right\rbrace$. We used the lowest literature value of $Q_\mathrm{c} (\theta =1) = Q_\mathrm{c,0} = 0.3$~eV \cite{Hodille_2024,Tamm1971}. The values of $a_\mathrm{m}$, $ \Delta Q_\mathrm{c}$ and $\gamma$ were adjusted to describe best the high-temperature tails in the TDS spectra.

To describe the rising edge of the spectra, we adjusted the activation energy for the bulk-to-surface transition to $E_\mathrm{BS} =0.085$~eV, which is considerably lower than the common assumption of $E_\mathrm{BS} = E_\mathrm{D}= 0.265$~eV. As was already mentioned, the position of the rising edge depends on the activation energy for the surface-to-bulk transition $E_\mathrm{S} + E_\mathrm{BS} + Q_\mathrm{c}$. Hence, we could as well reduce $Q_\mathrm{c,0}$ and make $E_\mathrm{BS} = E_\mathrm{D}$. However, too low values of $Q_\mathrm{c,0}$ seem unphysical, while low values of $E_\mathrm{BS}$ have been observed in DFT calculations \cite{Ajmalghan_2019,Ferro_2023}.

As can be seen from Fig.~\ref{ris:DP-exp-TESSIM-comp}, using the void parameters derived from TEM does not allow to match perfectly the D concentrations measured by NRA: for the 0.1~dpa sample the simulated D concentration is overestimated while for the 2.3~dpa sample it is underestimated. Therefore, in our simulations the width of the void-containing zone $\ell_\mathrm{v}$ increases with increasing damage dose. According to TEM observations of the 2.3~dpa sample, the voids are homogeneously distributed up to a depth of 2.5~$\mu$m (Fig.~\ref{fig:TEM-depth-res}), while the measured D depth distributions are less homogeneous on that depth scale. We used $\ell_\mathrm{v} = 2.5$~$\mu$m for the 2.3~dpa sample. Nevertheless, conisdering the limited void statistics from TEM, as well as the uncertanty of the void number density determination (related with the lammella thickness measurements), we consider the agreement between the experiments and the simulations to be reasonable. It is certainly possible to adjust the void parameters in the simulations to get a better agreement with the measured D depth distributions. 

Fig.~\ref{ris:TDS-exp-sim} shows the comparison of the experimental and the simulated TDS spectra. The model allows to reproduce reasonably the main desorption peak and the high-temperature shoulder. 
We also show the contributions of the $\rm D_2$ gas depletion in the void volume, chemisorbed D depletion at the void surface, and the bulk trap to the total TDS spectrum. The peak of the $\rm D_2$ gas depletion is located near 600~K. This temperature is frequently attributed to the D release from vacancies in W \cite{Hou2019,Zibrov2016}. This shows that attribution of the TDS peak position to the D release from a certain defect type is ambiguous.

The simulated TDS spectra exhibit a sharp low-temperature peak that is not visible on the experimental spectra. This may be related to a more complicated D adsorption/desorption kinetics at high gas pressures/surface coverages than in the present model. The the uncertainty of the equation of state (EOS) for the $\rm D_2$ gas at high pressures may also play a role \cite{hagelstein2015}. Lastly, our model approximates the actual size distribution of voids by using one void type with the corresponding average values.

The maximum calculated $\rm D_2$ gas pressure in the voids at 370~K is 5.68~GPa and the fugacity is $4.9 \times 10^{8}$~GPa, corresponding to the equilibrium with the maximum interstitial D concentration according to Sieverts' law (Eq.~\ref{eq:Sieverts} in the Appendix~\ref{app:model-descr}). This corresponds to the D density in the void volume of $1.48 \times 10^{29}$~$\text{D} / \text{m}^3$. In the case of the 2.3~dpa sample this corresponds to 20 874 $\rm D_2$ molecules in the void volume and 7051 chemisorbed D atoms at the void surface. The enormous difference between the pressure and the fugacity highlights how significant is the non-ideal gas behaviour at such pressures. 

It is known that D forms a molecular solid at high pressures \cite{10.1063/5.0002104}. The solid D melts at 370~K at a pressure near 7.5~GPa, while at 295~K already at about 5~GPa \cite{solid-h}. Hence, the $\rm D_2$ gas in the cavities could solidify during the sample cool down to room temperature after the plasma exposure. Although the $\rm D_2$ equation of state by Joubert \cite{Joubert2010,Joubert2011} used in the present work is also valid for solid D, explicit simulation of the phase transition represents an addition challenge. It is also worth mentioning that the $\rm D_2$ gas pressure in the voids is still below the minimum pressure required for the void growth via dislocation loop punching mechanism ($>10$~GPa for the present void sizes) \cite{KOLASINSKI2011S676}. 

\section{Summary}
In order to simulate the displacement damage produced by fusion neutrons, recrystallized tungsten (W) samples were irradiated by 20~MeV self-ions at 1350~K. The experiments covered a wide range of peak damage doses ranging from 0.001 to 2.3~dpa. To decorate the introduced defects with deuterium (D) without creating any additional damage, the samples were exposed to a low-temperature D plasma at 370~K. Quantitative absolute measurements of the trapped D concentrations in the samples were performed using $\rm ^{3}He$ nuclear reaction analysis. Since the irradiation-induced vacancy-type defects trap D, the measured local D concentrations are proportional to the local defect density. Except for the different irradiation temperature, all the experimental steps were essentially the same as in our previous irradiations at 290~K and 800~K \cite{Schwarz-Selinger_2023,tms-800K-data,SCHWARZSELINGER2017683,SCHWARZSELINGER2018228}, enabling a quantitative comparison.

In the cases of irradiation at 290~K and 800~K, the trapped D concentration increases with increasing damage dose up to 0.1~dpa, where it reaches a saturation value (1.8~at.\% at 290~K and 0.45~at.\% at 800~K) \cite{tms-800K-data}. However, in the case of irradiation at 1350~K, no clear trend towards saturation is visible: at 2.3~dpa the D concentration reaches 1.7~at.\%. Thermal desorption spectra (TDS) from the samples irradiated at 1350~K differed significantly from the spectra of the samples irradiated at 290~K and 800~K, demonstrating the change of the D trapping mechanism. Transmission electron microscopy (TEM) investigations of the samples irradiated to 0.1, 0.5, and 2.3~dpa revealed the presence of nm-sized irradiation-induced voids. The damage dose dependence of the void swelling correlated with the damage dose dependence of the trapped D concentration in these samples. Therefore, voids presumably play a dominant role in D trapping in the samples irradiated at 1350~K. Reaction-diffusion simulations assuming that all trapped D is located in the voids as $\rm D_2$ gas in the volume and as D atoms at the  surface could reasonably describe the measured TDS spectra.

The trapped D concentrations observed in the samples irradiated at 1350~K to 0.1--2.3~dpa are considerably higher compared with the extrapolation of the existing data on ion-irradiated tungsten at different temperatures (up to 1243~K). We attribute this to a non-monotonic temperature dependence of void swelling. Future studies should explore higher damage dosed (beyond 2.3~dpa) at 1350~K. Additional studies are also needed to explore even higher irradiation temperatures to further study the correlation between the void swelling and D retention.

\section{Acknowledgements}
The technical assistance of T.~Dürbeck, K.~Hunger, J.~Dorner, and M.~Fußeder is gratefully acknowledged. 
This work has been carried out within the framework of the EUROfusion Consortium, funded by the European Union via the Euratom Research and Training Programme (Grant Agreement No 101052200 — EUROfusion). Views and opinions expressed are however those of the author(s) only and do not necessarily reflect those of the European Union or the European Commission. Neither the European Union nor the European Commission can be held responsible for them.

\section{Appendix: Reaction-diffusion model of hydrogen isotope trapping and release from spherical voids}\label{app:model-descr}
\begin{figure}[!htb]
\center{\includegraphics[width=0.95\textwidth]{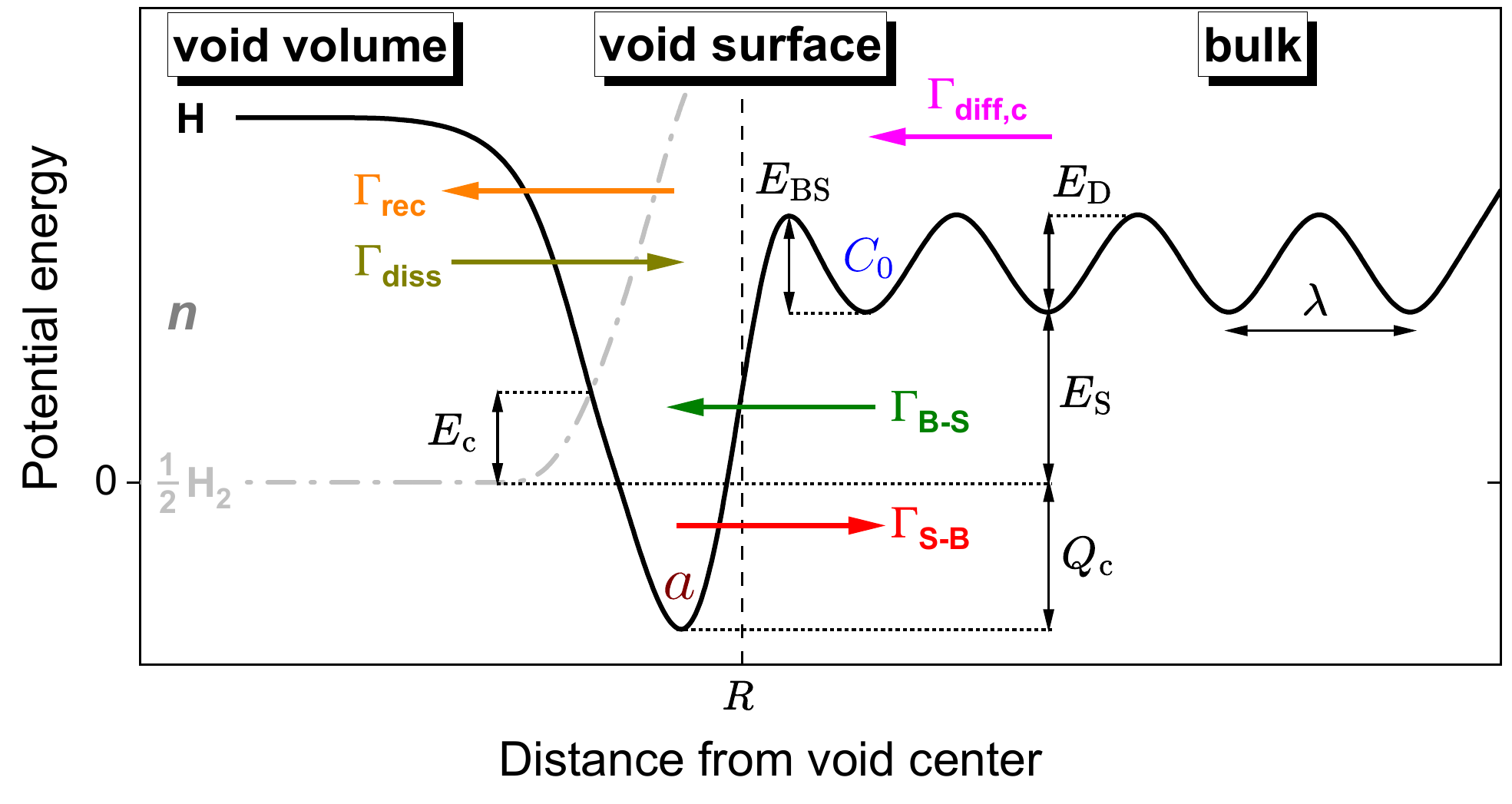}}
\caption{Schematic potential energy diagram for H atoms (solid line) and $\rm H_2$ molecules (dash-dotted line) near a void surface in a metal, where $E_\mathrm{c}$ is the activation energy for $\rm H_2$ dissociation, $Q_\mathrm{c}$ is the heat of chemisorption, $E_\mathrm{S}$ is the heat of solution, $E_\mathrm{BS}$ is the activation energy for the transition from the subsurface interstitial sites to the chemisorption sites, $E_\mathrm{D}$ is the activation energy for diffusion, $n$ is the density of $\rm H_2$ molecules, $a$ is the surface concentration of chemisorbed H atoms, $C_0$ is the concentration of interstitial H just outside the cavity surface, and $\lambda$ is the jump distance between the nearest interstitial sites. The indicated fluxes are explained in the text.}
\label{ris:potent-diagr}
\end{figure}

A metal has spherical voids with the radius $R$ and low number density $N_\mathrm{v} \ll R^{-3}$. We assume that H can be present as $\rm H_2$ gas in a void volume and as chemisorbed H atoms at a void surface \cite{Hou2019}. Their corresponding potential energy diagrams near the void surface are shown schematically in Fig.~\ref{ris:potent-diagr}. For each void we consider the concentration of interstitial H just outside the void surface $C_0$ ($\text{at.} / \text{m}^{3}$), the surface concentration of chemisorbed H atoms $a$ ($\text{at.}/ \text{m}^{2}$), and the density of $\rm H_2$ molecules in the void volume $n$ ($\text{molec.} /\text{m}^{3}$). The latter quantity is linked with the molar volume of the $\rm H_2$ gas $V_m$ ($\text{m}^3/\text{mole}$) as $n = N_A / V_m$, where $N_A$ is the Avogadro constant. Using the Joubert's equation of state \cite{Joubert2010,Joubert2011}
\begin{equation}\label{eq:eos-joubert}
V_\mathrm{m} (p,T) = \frac{\mathcal{R}T}{p} + \zeta + \sum_{i=1}^{5}\alpha_i \exp\left(- \frac{p}{\eta_i} \right)
\end{equation}
we can compute the corresponding $\rm H_2$ gas pressure $p$ (Pa) and the fugacity $f$ (Pa) in the void volume
\begin{equation}\label{eq:fugacity-Joubert}
\ln  \frac{f}{p} = \frac{1}{\mathcal{R}T} \left( \zeta p + \sum_{i=1}^{5} \alpha_i\eta_i \left[1-\exp\left(- \frac{p}{\eta_i} \right)\right] \right),
\end{equation}
where $T$ is the absolute temperature, $\mathcal{R}$ is the universal gas constant, $\alpha_1 = 4.86 \times 10^{-6}$, $\alpha_2 = 5.46 \times 10^{-6}$, $\alpha_3 = 4.342 \times 10^{-6}$, $\alpha_4 = -0.94 \times 10^{-6}$, $\alpha_5 = -1.79 \times 10^{-6}$, $\eta_1 = 5.35 \times 10^{8}$, $\eta_2 = 4.21 \times 10^{9}$, $\eta_3 = 3.99 \times 10^{10}$, $\eta_4 = 2.9 \times 10^{7}$, $\eta_5 = 8.02 \times 10^{7}$, $\zeta = 2.434 \times 10^{-6}$.

The fluxes  ($\text{at.}/( \text{m}^{2} \text{s})$) shown in Fig.~\ref{ris:potent-diagr} are defined as follows:
\begin{itemize}
\item Dissociative adsorption of $\rm H_2$ molecules:
\begin{equation}\label{eq:adsortion-flux}
\Gamma_\mathrm{diss} =  2 \frac{f}{\sqrt{2 \pi m k_\mathrm{B} T}} s_0 \exp \left(- \frac{2E_\mathrm{c}}{k_\mathrm{B} T} \right) \left(1 - \frac{a}{a_\mathrm{m}} \right)^2,
\end{equation}
where $m$ is the mass of a $\rm H_2$ molecule, $k_\mathrm{B}$ is the Boltzmann constant, $s_0$ is the initial sticking coefficient of $\rm H_2$ molecules on a bare surface, $E_\mathrm{c}$ is the activation energy for $\rm H_2$ dissociation, and $a_\mathrm{m}$ is the maximal possible surface concentration of chemisorbed H atoms.
\item Recombinative desorption of $\rm H_2$ molecules:
\begin{equation}\label{eq:recomb-flux}
\Gamma_\mathrm{rec} = 2 \chi_\mathrm{rec}  \exp \left(- \frac{2(E_\mathrm{c} + Q_\mathrm{c})}{k_\mathrm{B} T} \right) a^2,
\end{equation}
where $\chi_\mathrm{rec}$ is the recombination rate constant and $Q_\mathrm{c}$ is the heat of H chemisorption.
\item Transition from the subsurface interstitial sites to the chemisorption sites:
\begin{equation}\label{eq:bulk-surf-flux}
\Gamma_\mathrm{B-S} = \nu_\mathrm{BS} \exp \left(- \frac{E_\mathrm{BS}}{k_\mathrm{B} T} \right)  \left(1 - \frac{a}{a_\mathrm{m}} \right) \lambda C_0,
\end{equation}
where $E_\mathrm{BS}$ and $\nu_\mathrm{BS}$ are the activation energy and the attempt frequency for the bulk-to-surface transition, and $\lambda$ is the jump distance between the nearest interstitial sites. 
\item Transition from the chemisorption sites to the subsurface interstitial sites:
\begin{equation}\label{eq:surf-bulk-flux}
\Gamma_\mathrm{S-B} = \nu_\mathrm{SB} \exp \left(- \frac{E_\mathrm{S} + E_\mathrm{BS} + Q_\mathrm{c}}{k_\mathrm{B} T} \right)  a,
\end{equation}
where $\nu_\mathrm{SB}$ is the attempt frequency for the surface-to-bulk transition and $E_\mathrm{S}$ is the heat of H solution.
\item Diffusion towards the cavity surface:
\begin{equation}\label{eq:analyt-diff-flux}
\Gamma_\mathrm{diff,v} = \frac{D(T)}{R} \left[C-C_0\right],
\end{equation}
where $C$ is the interstitial H concentration far away from the cavity described by Eq.~\ref{eq:3dfick-second}, $D(T) = D_0 \exp \left(- \frac{E_\mathrm{D}}{k_\mathrm{B} T} \right)$ is the H diffusivity, and $E_\mathrm{D}$ is the activation energy for diffusion.
\end{itemize}

The material conservation equations are given by:
\begin{equation}\label{eq:C0-vs-t}
\lambda \frac{\partial C_0}{\partial t} = \Gamma_\mathrm{S-B} - \Gamma_\mathrm{B-S} + \Gamma_\mathrm{diff,v},
\end{equation}
\begin{equation}\label{eq:a-vs-t}
\frac{\partial a}{\partial t} = \Gamma_\mathrm{diss} - \Gamma_\mathrm{rec} - \Gamma_\mathrm{S-B} + \Gamma_\mathrm{B-S},
\end{equation}
\begin{equation}\label{eq:mat-balance}
2V\frac{\partial n}{\partial t} = A \left( \Gamma_\mathrm{rec} - \Gamma_\mathrm{diss} \right),
\end{equation}
where $V$ is the cavity volume, and $A$ is the cavity surface area.

Under the thermodynamic equilibrium all time derivatives vanish and $\Gamma_\mathrm{diff,v}= 0$, yielding the Sieverts' law \cite{MARCHI2007100}:
\begin{align}
C = C_0 = S(T) \sqrt{f}, \label{eq:Sieverts} \\
S(T) = S_0 \exp \left(- \frac{E_\mathrm{S}}{k_\mathrm{B} T} \right). \label{eq:S-def}
\end{align}

This links the rate coefficients with the solubility constant: 
\begin{equation}\label{eq:S0-prefac}
S_0 = \frac{1}{\sqrt[4]{2 \pi m k_\mathrm{B} T}} \sqrt{\frac{s_0}{\chi_\mathrm{rec}}} \frac{\nu_\mathrm{SB}}{\lambda \nu_\mathrm{BS}},
\end{equation}
which allows us to compute $\nu_\mathrm{SB}$.
We take $\nu_\mathrm{BS}$ equal to the attempt frequency of interstitial diffusion \cite{Takagi}:
\begin{equation}\label{eq:D0-prefac}
D_0 = \gamma \frac{\nu_\mathrm{BS} \lambda^2}{6},
\end{equation}
where $\gamma$ is the number of nearest neighbour interstitial sites. For tetrahedral sites in bcc W $\gamma = 4$.

The macroscopic evolution of the concentration of interstitial H far away from the cavity $C(x,t)$ ($\text{at.} / \text{m}^{3}$) is governed by the diffusion and the trapping/release from voids and other traps:
\begin{equation}\label{eq:3dfick-second}
\frac{\partial C(x,t)}{\partial t} = D(T) \frac{\partial^2 C(x,t)}{\partial x^2} -A \frac{D(T)}{R} N_\mathrm{v} (x) \left[C(x,t)-C_0(x,t)\right] - \frac{\partial C_\mathrm{t}}{\partial t}.
\end{equation}
In the conventional trapping model (McNabb and Foster), which assumes one H atom per trap, the trapping/release term is given by \cite{Takagi}:
\begin{equation}\label{eq:trap-eq-classical}
\frac{\partial C_\mathrm{t}}{\partial t} = \nu_\mathrm{tr} \exp \left(- \frac{E_\mathrm{tr}}{k_\mathrm{B} T} \right) \frac{\left( N_\mathrm{t} - C_\mathrm{t} \right)}{h \rho} C - \nu_\mathrm{dt} \exp \left(- \frac{E_\mathrm{dt}}{k_\mathrm{B} T} \right) C_\mathrm{t}, 
\end{equation}
where $C_\mathrm{t}$ is the concentration of trapped H ($\text{at.} / \text{m}^{3}$), $N_\mathrm{t}$ is the trap density ($1/ \text{m}^{3}$), $E_\mathrm{tr}$ and $E_\mathrm{dt}$ are the activation energies for trapping and detrapping, respectively, $\nu_\mathrm{tr}$ and $\nu_\mathrm{dt}$ are the corresponding attempt frequencies, and $h$ is the number of interstitial sites per lattice atom.

The system of these coupled equations is solved numerically using the NDSolve in Wolfram Mathematica. The parameters used in the present simulations are listed in Tab.~\ref{tabl:sim-param}. 

\begin{table}[!h]
\label{tabl:sim-param}
\begin{center}
\begin{tabular}{|p{0.4\textwidth}|l|l|}
\hline
W density & $\rho$ & $6.306 \times 10^{28}$~$\text{at.}/\text{m}^3$ \\
Jump distance between nearest interstitial sites & $\lambda$ & $1.12 \times 10^{-10}$~m \\
Number of interstitial sites per lattice atom & $h$ & 6 \\
Diffusivity & D(T) & $1.5 \times 10^{-7} \exp \left(- \frac{0.265 \; \text{eV}}{k_\mathrm{B} T} \right) \; \text{m}^2 / \text{s}$ \cite{holzner-phd} \\
\hline
%Solubility & S(T) & $\frac{8.6}{4.5} \times 10^{-5} \exp \left(- \frac{0.92 \; \text{eV}}{k_\mathrm{B} T} \right) \; \text{at. frac.} / \text{Pa}^{1/2}$ \\
Activation energy for bulk-to-surface transition & $E_\mathrm{BS}$ & $0.085$~eV \\
Attempt frequency for bulk-to-surface transition & $\nu_\mathrm{BS}$ & $1.8 \times 10^{13}$~$1/ \text{s}$ \\
Heat of solution & $E_\mathrm{S}$ & 1.0~eV \\
Attempt frequency for surface-to-bulk transition & $\nu_\mathrm{SB}$ & $9.5 \sqrt[4]{T} \times 10^{12}$~$1 / \text{s}$ \\
Heat of chemisorption & $Q_\mathrm{c}$ & $0.3 + 1.35 \left\lbrace 1 - \exp \left[- 0.5 \left(1 - \frac{a}{a_\mathrm{m}} \right) \right] \right\rbrace$ eV \\
Recombination rate constant & $\chi_\mathrm{rec}$ & $10^{-6}$~$\text{m}^2 / \text{s}$ \cite{Tamm1971} \\
Activation energy for $\rm H_2$ dissociation &$E_\mathrm{c}$ & 0  \\
Sticking coefficient of $\rm H_2$ molecules on a H-free surface & $s_0$ & 1 \\
Maximal surface concentration of chemisorbed H & $a_\mathrm{m}$ & $3.5 \times 10^{19}$~$\text{at.}/\text{m}^2$ \\
\hline
Trap density & $N_\mathrm{t}$ & $1.73 \times 10^{-5}$~at.~frac. \\
Activation energy for trapping & $E_\mathrm{tr}$ & $0.265$~eV \\
Attempt frequency for trapping & $\nu_\mathrm{tr}$ & $1.2 \times 10^{13}$~$1/ \text{s}$ \\
Detrapping energy & $E_\mathrm{dt}$ & $1.18$~eV \\
Detrapping attempt frequency & $\nu_\mathrm{dt}$ & $1 \times 10^{13}$~$1/ \text{s}$ \\
\hline
\end{tabular}
\end{center}
\caption{The simulation parameters.}
\end{table}

\bibliographystyle{elsarticle-num} 
\bibliography{1350K-bibl}

\end{document}